\pdfoutput=1 
\documentclass{article}
\usepackage{graphicx}
\usepackage{float}
\usepackage{vmargin}
\usepackage{authblk}
\usepackage{amsmath}
\usepackage{amsfonts}
\usepackage{amssymb}
\usepackage{relsize}
\usepackage{bbm}
\usepackage{bm}
\usepackage{hyperref}
\usepackage{cleveref}
\usepackage{multirow}
\usepackage{lipsum}
\usepackage{cancel}

\usepackage{tikz}
\usetikzlibrary{arrows}
\usetikzlibrary{trees}
\usetikzlibrary{decorations.markings}
\usetikzlibrary{arrows.meta}

\tikzset{point/.style={insert path={ node[scale=6*sqrt(\pgflinewidth)]{.} }}}

\def\E{\mathbb{E}}

\definecolor{HLcolor}{rgb}{0,0.0,0}

\title{Regional and spatial dependence of poverty factors in Thailand, and its use into Bayesian hierarchical regression analysis}
\author[1,*]{Irving Gómez-Méndez}
\author[2]{Chainarong Amornbunchornvej}
\affil[1]{CMKL University}
\affil[*]{Corresponding author: \href{gomendez.irving@gmail.com}{gomendez.irving@gmail.com}}
\affil[2]{National Electronics and Computer Technology Center (NECTEC), \href{chainarong.amo@nectec.or.th}{chainarong.amo@nectec.or.th}}
\date{}

\begin{document}

\maketitle



\section*{Highlights}

\begin{itemize}
    \item We proposed inferred boundaries of regions of Thailand that can explain better the poverty dynamics, instead of the usual government administrative regions.
    \item The proposed regions maximize a trade-off between poverty-related features and geographical coherence.
    \item Northern, Northeastern Thailand, and in less extend Northcentral Thailand are the regions that require more attention in the aspect of poverty issues.
    \item The use of principal components and Moran's clusters can help to identify regions that need more attention and to establish the high-priority aspects to mitigate.
    \item Bangkok-Pattaya is the only region whose average years of education is above 12 years. The national average years of education is 10.87 years.
\end{itemize}

\begin{abstract}
  Poverty is a serious issue that harms humanity progression. The simplest solution is to use one-shirt-size policy to alleviate it. Nevertheless, each region has its unique issues, which require a unique solution to solve them. In the aspect of spatial analysis, neighbor regions can  provide useful information to analyze issues of a given region. In this work, we proposed inferred boundaries of regions of Thailand that can explain better the poverty dynamics, instead of the usual government administrative regions.  The proposed regions maximize a trade-off between poverty-related features and geographical coherence.  We use a spatial analysis together with Moran's cluster algorithms and Bayesian hierarchical regression models, with the potential of assist the implementation of the right policy to alleviate the poverty phenomenon. We found that all variables considered show a positive spatial autocorrelation. The results of analysis illustrate that 1) Northern, Northeastern Thailand, and in less extend Northcentral Thailand are the regions that require more attention in the aspect of poverty issues, 2) Northcentral, Northeastern, Northern and Southern Thailand present dramatically low levels of education, income and amount of savings contrasted with large cities such as Bangkok-Pattaya and Central Thailand, and 3) Bangkok-Pattaya is the only region whose average years of education is above 12 years, which corresponds (approx.) with a complete senior high school.
\end{abstract}

\section{Introduction}

\subsection{Motivation}
Poverty is a serious issue in many developing countries. In 2030, one of the United Nation (UN)'s plans is to eliminate poverty worldwide~\cite{zhou2022geography}. Thus, finding optima policies are necessary to help the governments cope with the problem~\cite{AMORNBUNCHORNVEJ2023e15947,zhang2023alleviating,okpala2023socio}. 

The simplest solution to combat poverty for government is to have a single one-shirt-size policy to solve each aspect of poverty issues~\cite{gomez2024income}. However, not all regions share similar nature of challenges, since each region typically has its own unique conditions, which requires a custom-made solution to solve regional problems~\cite{berdegue2002rural,commins2004poverty,pringle2000cross,10.1145/3424670}. For instance, focusing on lacking of health care support, it is possible for some areas to build a new hospital while it might be harder for remote areas in mountain regions to build it. Hence, instead of insisting on having a new hospital, providing telehealth~\cite{blandford2020opportunities} might possibly alleviate this issue in remote places. While one-shirt-size policy is not an optimal solution to solve poverty issues, having custom-made policies for every region also faces the challenge of limitation of resource~\cite{10.1145/3424670}.

On the other hand, some regions might share similar problems with neighbor regions, creating regional clusters~\cite{10.1145/3424670}. Hence, instead of trying to make a  custom-made policy for each region, in this work, we focus on exploring groups of regions that share similar properties so that a single policy might be custom-made developing to solve the regional unique issues. The results are the inferred group of regions that share similar properties. These inferred groups of regions can be used for developing a custom-made policy to precisely and effectively combat regional poverty issues. For instance, we use these groups to explore the relationship between income and education at regional and national levels using Bayesian hierarchical regression.

\subsection{Related works}
Poverty is conceived as a multidimensional problem beyond only a monetary aspect~\cite{zhou2022geography,alkire2022revising,10.1145/3424670,gomez2024income}. This complexity of poverty creates several challenges in order to alleviate it~\cite{zhou2022geography}. Currently, a widely used statistic to measure poverty is the  Multidimensional Poverty index (MPI)~\cite{alkire2021global,alkire2015multidimensional,alkire2022revising}. This index covers several dimensions that are related to poverty such as lack of education, health issues, lack of resource, etc.     

Even though MPI illustrates the complexity of poverty issues well, we cannot tell whether we need a one-shirt-size policy or custom-made policies for a given region or a group of sub-regions~\cite{gomez2024income,10.1145/3424670} using just such index. In other words,  the MPI cannot indicate whether each region has unique issues or share similar nature of problems with its neighbors.

With this challenge is mind, the work in~\cite{10.1145/3424670} utilized Minimum Description Length (MDL) and Gaussian Mixture Models~\cite{grun2007applications,grun2006fitting,JSSv011i08} to focus on the multi-resolution aspects of poverty in the partitioning inference problem; whether we need a one-shirt-size policy for a specific resolution of regions.
The method proposed in~\cite{10.1145/3424670} tried to find correct resolution of partitions of provinces or sub-regions inside the provinces where a single policy can be placed. However, this approach was unable to analyzed the details of spatial dependency between larger partitions (e.g. provinces) and its sub-regions into the model.

The work presented in~\cite{gomez2024income} attempted to fill this gap by analysing the spatial hierarchical structure of partitions of regions using Bayesian's statistics and modeling, which was deployed by government policy makers to develop policies~\cite{fienberg2011bayesian,finucane2018works}. The same work also found that there were common problems and dependencies among lager partitions to the sub-region within them (e.g. almost all provinces in the same region have the issue of lacking of education). Nevertheless, the gap in the field still remains, there is no work focusing on regional and spatial dependence of poverty of Thailand; whether a given region shares similar issues among its neighbors. 

\subsection{Contribution}
As pointed out by ~\cite{gomez2024income}, one-shirt-size policy cannot handle poverty issues well since each province has its unique challenges, while having a custom-made policy for each province separately is unrealistic due to limitation of resources as well as ignoring dependencies of characteristics between different provinces. Thus, it results more convenient to present policies for regions that incorporate similar provinces, presenting an optimum solution to the two extremes of one-shirt-size and custom-made policies. In this work, we present our approach using spatial statistical analysis and Bayesian hierarchical regression models~\cite{gelman2006data,congdon2019bayesian}, and related statistics to find geographically consistent clusters for the provinces of Thailand, which can represent better the challenges affronted by the households considering the differences between the regions. We found that 1) all variables considered show a positive spatial autocorrelation, 2) a spatial analysis together with cluster algorithms can help to identify specific issues that affect each region, 3) the use of principal components and Moran's clusters~\cite{https://doi.org/10.1111/j.1538-4632.2007.00708.x} can help to identify regions that need more attention and to establish the high-priority aspects to mitigate, and 4) Bayesian hierarchical regression can describe the relation between income and education at a regional and national levels. 

The rest of the article is organized as follows.  In Section~\ref{sec:Method}, the details of material and methods are provided.  In Section~\ref{sec:Res}, the results are illustrated. We discuss the results in Section~\ref{sec:Discussion} observed throughout the work and we also present the conclusions of this work.

\section{Material and methods}
\label{sec:Method}
\subsection{Data}

The dataset used throughout this work has been collected in 2022 by Thai government agencies. The main purpose of this dataset is to support poverty alleviation policy making in the Thai People Map and Analytics Platform project (\href{https://www.tpmap.in.th/about_en}{www.TPMAP.in.th})~\cite{10.1145/3424670}. The dataset has 12,983,145 observations, each one corresponding to a household. The number of households consulted for each province goes from 44,012 up to 645,433.

\subsection{Global Moran's I and Moran's plot}

To have a perspective of possible regions and the spatial structure of a variable, we can apply a classification technique, grouping the provinces into mutually exclusive and exhaustive groups. For this purpose we can consider the Fisher-Jenks algorithm ~\cite{jenks1967data}, which has been extensively used in spatial analyses ~\cite{rey2023geographic,zongfan2022spatiotemporal,hou2022improved,khamis2018segmentation,chen2013research}. This algorithm optimizes the arrangement of clusters, reducing the within classes variance while maximizing the variance between classes. However, since the Fisher-Jenks algorithm does not consider the geographical structure, it can potentially create clusters that are spatially disconnected.

To identify clusters that incorporate the spatial structure of the data, we need to construct a network that reflects the geographical connection between the provinces of Thailand, creating a topology that resembles the country. A usual approach to build such network is to connect each province $i$ with its $k$-nearest neighbors, $\text{knn}(i)$ ~\cite{fotheringham2003geographically}, inducing weights of the form \[w_{ij}=\begin{cases}1/k &\text{if province }j\in \text{knn}(i), \\ 0 &\text{otherwise}.\end{cases}\] Using these spatial weights, we can calculate the spatial lag of a variable for province $i$, defined as \[\text{lag-}y_i=\sum_{j=1}^n w_{ij}y_j,\] where $y_i$ is the observed value of the attribute for the province $i$.

To determine if there is statistical evidence of spatial structure for the attribute, we can estimate its spatial autocorrelation. When the variable is geographically distributed such that high values are nearby other high values and low values are nearby other low values, we say that the variable has a positive spatial autocorrelation. On the other hand, when the attribute is distributed such that high and low values are close, we say that the variable presents a negative spatial autocorrelation. Probably, the most commonly used statistic to estimate the spatial autocorrelation of a variable is Moran’s I ~\cite{moran1948interpretation} (see for example \cite{mendez2021regional,mendez2020convergence}), which is given by \[I=\frac{\sum_{i=1}^n\sum_{j=1}^n w_{ij}z_iz_j/S_0}{\sum_{\ell=1}^n z_\ell^2/n},\] where $z_i=y_i-\bar{y}$, and $S_0=\sum_{i=1}^n\sum_{j=1}^n w_{ij}$. It can be proved that, under absence of spatial autocorrelation, the expected value of Moran's I is \[\E(I)=-\frac{1}{n-1}.\]

On the other hand, since $\sum_{j=1}^n w_{ij}=1$, then $S_0=n$. And, Moran's I statistic can be rewritten as
\begin{align}
I &= \frac{\sum_{i=1}^n\sum_{j=1}^n w_{ij}z_iz_j}{\sum_{\ell=1}^n z_\ell^2} \nonumber\\\nonumber\\
&= \frac{\sum_{i=1}^n z_i\sum_{j=1}^n w_{ij}z_j}{\sum_{\ell=1}^n z_\ell^2}\nonumber\\\nonumber\\
&= \frac{\sum_{i=1}^n z_i\text{lag-}z_i}{\sum_{\ell=1}^n z_\ell^2},\label{eq:MoranI}
\end{align}
which corresponds with the least squares estimator of the slope in the linear regression \[\text{lag-}z=\beta_0+\beta_1 z + \varepsilon.\] Also note that, because $z_i=y_i-\bar{y}$, then $\hat{\beta}_0=0$.

Therefore, we can create a scatter plot in which the variable of interest is displayed against its spatial lag, whose estimated slope by least squares corresponds with its Moran's I. This graph is known as Moran's plot or Moran's scatter plot.

Since, by definition, the spatial lag is a weighted average of the $k$-nearest neighbors of each observation, then the top-right quadrant in Moran's plot presents observations with an attribute above the average whose neighbors' attribute is also above the average, and corresponding with a positive spatial autocorrelation. Thus, we name this quadrant as high-high (HH). Similarly, we can name the other quandrants as low-high (LH) for the top-left, low-low (LL) for the bottom-left, and high-low (HL) for the bottom right.

\subsection{Local Moran's I and Moran clusters}

Consider the expression of Moran's I statistic given by \Cref{eq:MoranI}, and note that it can be expressed as
\begin{align*}
I &= \frac{1}{n}\frac{\sum_{i=1}^n z_i\text{lag-}z_i}{\sum_{\ell=1}^n z_\ell^2/n} \\\\
&= \frac{1}{n} \sum_{i=1}^nI_i,
\end{align*}
where \[I_i=\frac{z_i\text{lag-}z_i}{\sum_{\ell=1}^n z_\ell^2/n}\] is known as the local Moran's I for the observation $i$.

Using Moran's plot and local Moran's I, we can create four clusters-- namely: HH, HL, LL, and LH--form by the observations whose local Moran's I is significantly different from its expected value under the hypothesis of no spatial correlation, depending on the quadrant where the observation is located in the Moran's plot. These clusters are usually named Moran clusters.

\subsection{Regionalization}

While Moran's clusters can detect and characterize regions of Thailand based on their poverty dynamics, they present two important drawbacks. First, it does not guarantee to find not fragmented clusters. Second, since the clusters are created only with provinces whose local Moran's I is significantly different from its expected value under the hypothesis of no spatial correlation, several provinces might not be assigned to any cluster. To ensure that clusters are not spatially fragmented and that every province is assigned to some cluster, we turn into regionalization.

Regionalization methods are clustering techniques that impose a spatial constraint on clusters. In other words, the result of a regionalization algorithm contains clusters with areas that are geographically coherent, in addition to having coherent data profiles. To consider the possibly non-linear interactions between the variables, we consider an agglomerative hierarchical clustering, considering the ward linkage and adding the spatial constraint given by the matrix of spatial weights.

To determine the number of regions, we have to take into account the geographic coherence of the regions as well as its feature coherence. For the geographic coherence we consider the isoperimetric quotient of the regions, $IPQ=4\pi A/L^2$, where $A$ is its area and $L$ its perimeter. According to the isoperimetric inequality, $IPQ\leq 1$, and $IPQ=1$ if and only if we consider a circumference, while non-convex shapes (``wormy'' shapes) would have a low $IPQ$. Thus, we try to achieve a high $IPQ$ for each region in order to achieve geographic coherence.

On the other hand, to measure the feature coherence of the regions we consider two metrics: the silhoutte score ~\cite{rousseeuw1987silhouettes} and the Calinski-Harabasz score ~\cite{calinski1974dendrite}. Both of them tried to maintain the feature coherence comparing the variance within the regions against the variance between them.

\subsection{Hierarchical and spatial regression}

Once we can group all provinces of Thailand into geographically coherent regions, we can implement a hierarchical Bayesian model over such regions as proposed in ~\cite{gomez2024income}, explaining some variable of interest in terms of other variables at a regional and national levels.

On the other hand, since we count with the spatial weights $w_{ij}$ we can also perform a geographically weighted regression model (GWR), which has been extensively used in spatial analyses in social and economic contexts  ~\cite{fotheringham2003geographically,li2019spatial,farahmand2014spatial}.
\section{Results}
\label{sec:Res}
\subsection{Moran clusters for each variable}

\begin{figure}
	\centering
	\includegraphics[width=\textwidth]{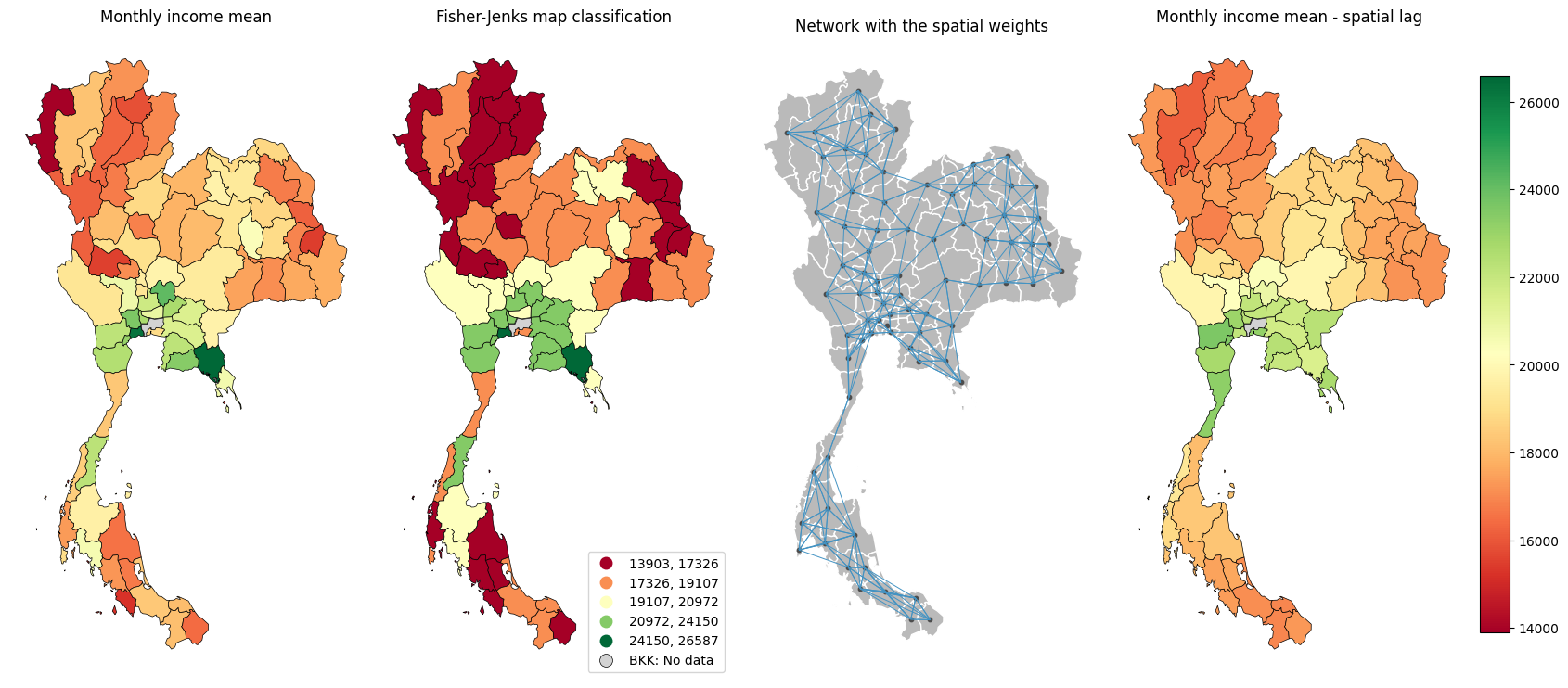}
	\caption{From left to right, the maps depicts the following information. First: We present the monthly average income reported per household in each province of Thailand. Second: We group the provinces into five different clusters using the Fisher-Jenks algorithm. Third: We present the topology induced by the spatial network, each province is connected with its five nearest neighbors. Fourth: We present the spatial-lag of the monthly average income.}
	\label{fig:SpatialNetwork}
\end{figure}

For its direct relation with poverty, we illustrate our approach analyzing the monthly average income of households.

On the first map of \Cref{fig:SpatialNetwork}, we present the observed average income per province. It is easy to observe that the area around Bangkok and Eastern Thailand reports the highest income. However, we cannot identify from this map if there are regions significantly different and, in case such regions exist, we cannot tell how they are composed. Thus, to have a better perspective of the possible regions and the spatial structure of the income, we can apply the Fisher-Jenks algorithm. The clusters identified through this technique are presented in the second map of \Cref{fig:SpatialNetwork}, where it becomes clear that provinces around Bangkok and Eastern Thailand belong to the two clusters with the highest reported income, while most of the provinces of Nothern, Northeastern and Southern Thailand belong to the two clusters with the lowest reported income. However, these clusters do not consider the spatial structure presented in the provinces and, as a consequence of such lack of spatial structure, all the clusters are geographically disconnected.

To identify clusters that incorporate the spatial structure of the data, we need to create a network that reflects the connection between the provinces of Thailand. We have seen in practice that considering the five nearest neighbors to build such network maintains the analysis local, while reproducing the spatial structure of Thailand, as it is depicted in the third map of \Cref{fig:SpatialNetwork}. Using these spatial weights, we get the spatial lag of the monthly income, which is presented in the fourth map of \Cref{fig:SpatialNetwork}. Since the spatial lag is, by definition, a weighted average, it creates a smoother transition of the income between the provinces.

\begin{figure}
	\centering
	\includegraphics[width=\textwidth]{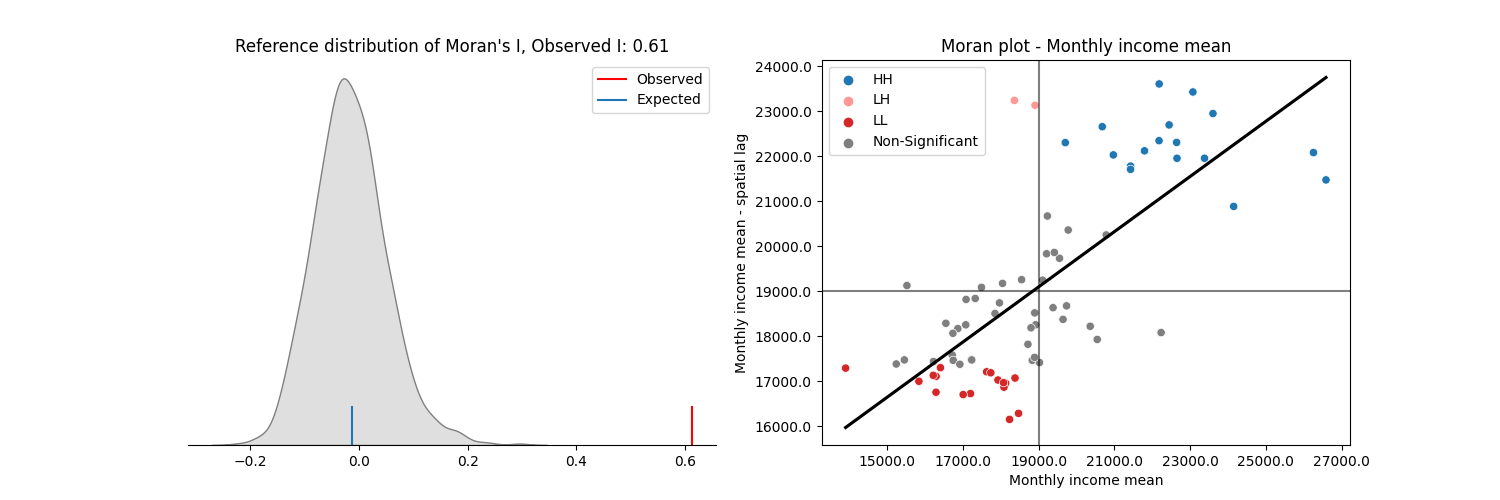}
	\caption{Left: Reference distribution for Moran's I in the absence of spatial autocorrelation for the monthly average income. Right: Moran's plot, the observations have been colored if their local Moran's I is significantly different from its expected value in the absence of spatial correlation}
	\label{fig:IncomeMoranPlot}
\end{figure}

Considering these spatial weights, we calculate Moran's I statistic to determine if there is statistical evidence of spatial structure for the income. To estimate the statistical significance of Moran's I, we permute the characteristic 1000 times, calculating the statistic for each permutation, which gives us a reference distribution in the absence of spatial structure. On the left side of \Cref{fig:IncomeMoranPlot} we present such reference distribution, we also show its expected value in absence of spatial structure, and the observed value of the statistic. Based on this analysis, we can conclude that there is evidence of spatial structure for the average monthly income. Furthermore, since the I statistic is positive, we can conclude that provinces with similar income tend to be close, creating possible clusters of provinces with similar income.

On the right side of \Cref{fig:IncomeMoranPlot}, we present Moran's scatter plot which presents the average monthly income for each province against its corresponding spatial lag. We have colored those observations where we rejected the absence of spatial autocorrelation considering their local Moran's I statistic and a significance level of 0.05.

\begin{figure}
	\centering
	\includegraphics[width=\textwidth]{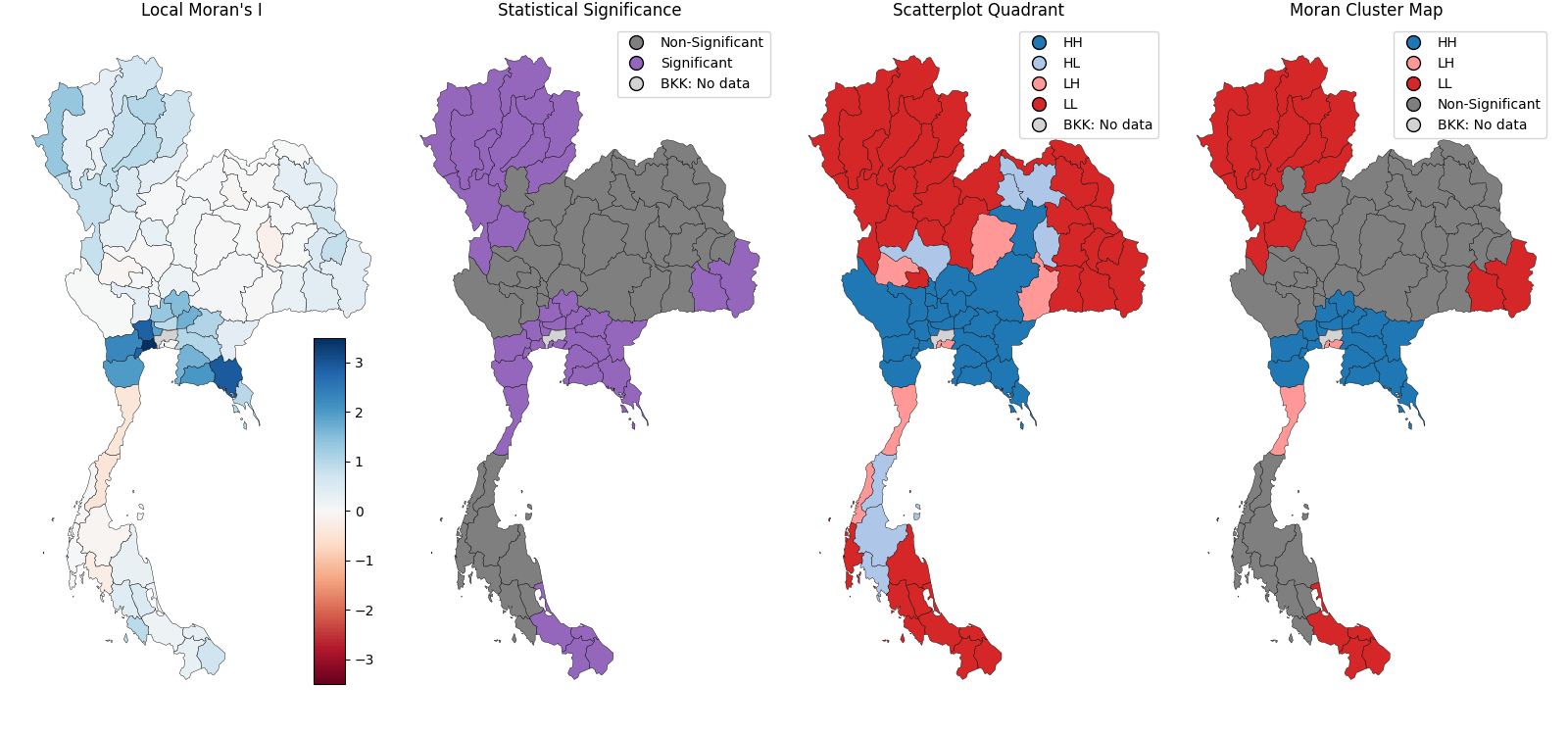}
	\caption{Clusters found using local Moran's I statistic. From left to right, the maps depicts the following information. First: We present the local Moran's I statistic for each province. Second: We present the provinces for which we reject the hypothesis of non-spatial autocorrelation. Third: We colored the provinces accordingly to the quadrant where they belong in Moran's plot. Fourth: We present Moran's clusters.}
	\label{fig:IncomeClusterMap}
\end{figure}

In \Cref{fig:IncomeClusterMap} we present the clusters found for the average monthly income using the local I statistics. From left to right, the maps present: the value of each local Moran's I statistic, those provinces whose local I is statistically different from its expected value under no spatial correlation assumption, the quadrant where each province belongs in the Moran's scatter plot, and Moran's clusters.

We can observe four well-distinguished clusters in the map. Around the area of Bangkok and Eastern Thailand we observe that the provinces form a cluster of high income. The other three clusters at Northern, Northeastern and Southern Thailand are clusters where the provinces present low income.

Due to the multidimensional nature of poverty, we reproduce this analysis for other eight variables, representing different aspects of the social problem, such as education, income, inequality, debt, and living aspects. The relation between the variables and the aspect that they represent is shown in \Cref{tab:VariablesAspects}. Four of these variables (percentage of households with no savings, percentage of households with formal debt, alcohol consumption, and smoking) where considered in ~\cite{gomez2024income}, corresponding with variables with the largest percentage of households that reported having an issue. To estimate the years of education, we follow the approach proposed in ~\cite{gomez2024income}, where the number of years of education is assigned accordingly to the rule presented in \Cref{tab:Education}.

\begin{table}
	\centering
	\begin{tabular}{|l|l|c|}
	\hline 
	\textbf{Variable} & \textbf{Aspect} & \textbf{Moran's I} \\ 
	\hline 
	Years of education & Education & 0.64 \\
	\hline 
	Monthly income & \multirow{3}{*}{Income} & 0.61 \\ 
	Yearly savings & & 0.25 \\
	Percentage of households without savings & & 0.56 \\
	\hline 
	Monthly income ratio 20:20 & \multirow{2}{*}{Inequality} & 0.47 \\
	Gini index & & 0.31 \\ 
	\hline 
	Percentage of households with formal debt & Debt & 0.72 \\
	\hline 
	Alcohol consumption & \multirow{2}{*}{Living aspect} & 0.53 \\
	Smoking & & 0.68 \\
	\hline 
	\end{tabular}
	\caption{Poverty factors, and their respective Moran's I statistic.}
	\label{tab:VariablesAspects}
\end{table}

\begin{table}
	\centering
	\begin{tabular}{|l|c|}
	\hline
	\textbf{Education} & \textbf{Years of education} \\
	\hline
	Uneducated & 0 \\ 
	\hline
	Kindergarten & 0 \\
	\hline
	Pre-elementary school & 3 \\
	\hline
	Elementary school & 6 \\
	\hline
	Junior high school & 9 \\
	\hline
	Senior high school & 12 \\
	\hline
	Vocational degree & 14 \\
	\hline
	Bachelor degree & 16 \\
	\hline
	Post-graduate & 19 \\
	\hline
	\end{tabular}
	\caption{Estimated years of education for each educational grade.}
	\label{tab:Education}
\end{table}

\begin{figure}
	\centering
	\begin{minipage}[c]{0.24\textwidth}
		\includegraphics[width=\textwidth]{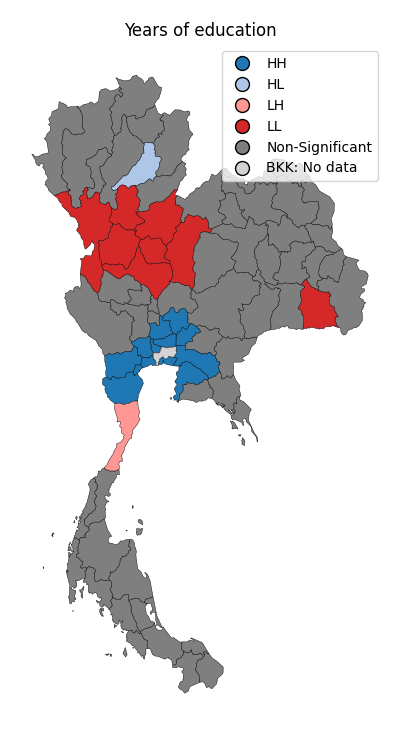}
	\end{minipage}
	\begin{minipage}[c]{0.24\textwidth}
		\includegraphics[width=\textwidth]{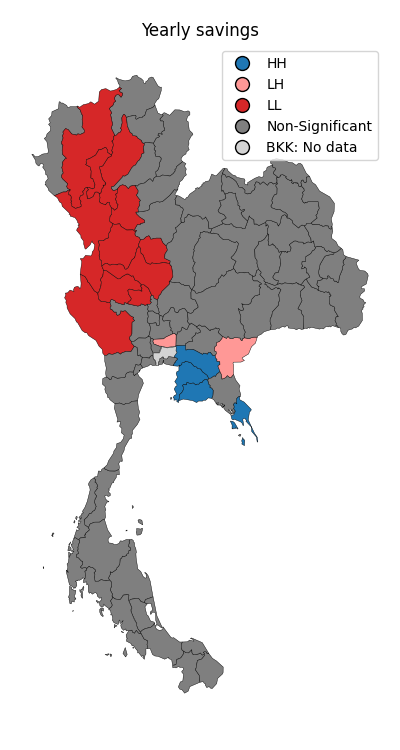}
	\end{minipage}
	\begin{minipage}[c]{0.24\textwidth}
		\includegraphics[width=\textwidth]{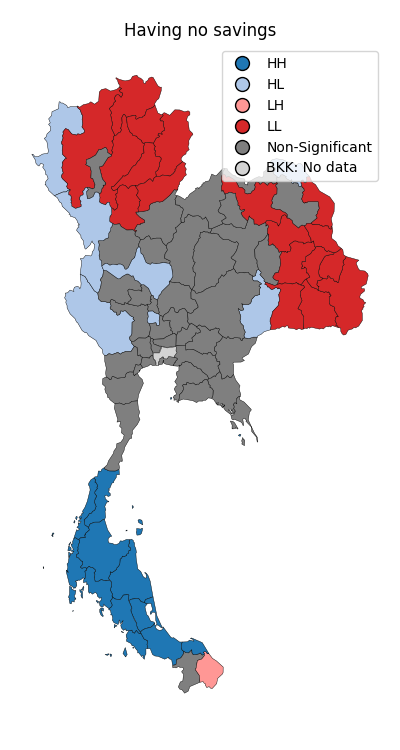}
	\end{minipage}
	\begin{minipage}[c]{0.24\textwidth}
		\includegraphics[width=\textwidth]{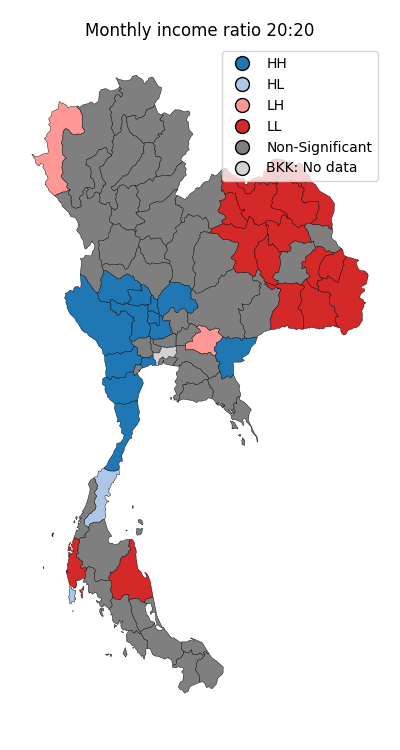}
	\end{minipage}
	\begin{minipage}[c]{0.24\textwidth}
		\includegraphics[width=\textwidth]{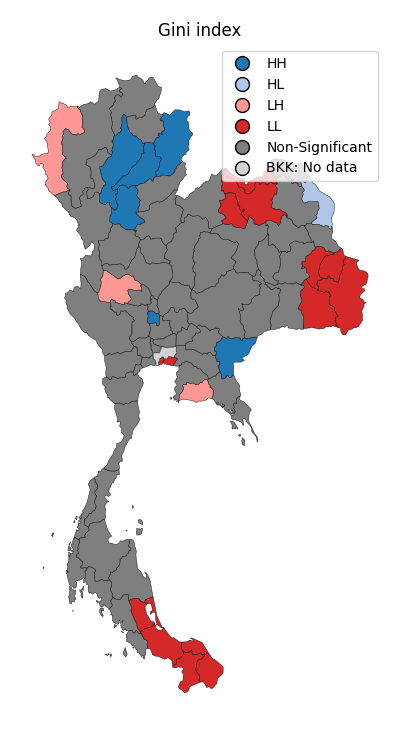}
	\end{minipage}
	\begin{minipage}[c]{0.24\textwidth}
		\includegraphics[width=\textwidth]{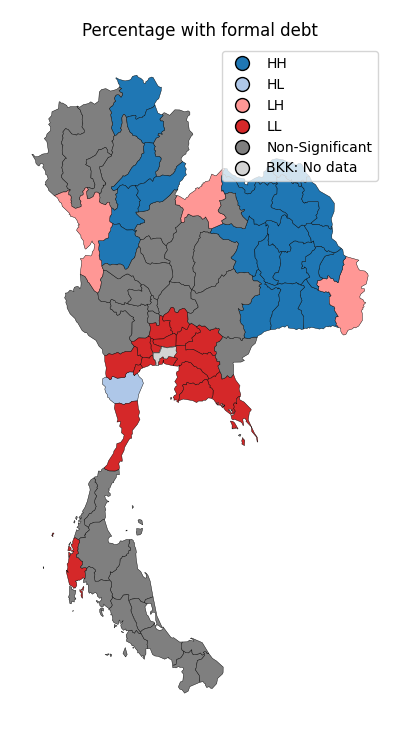}
	\end{minipage}
	\begin{minipage}[c]{0.24\textwidth}
		\includegraphics[width=\textwidth]{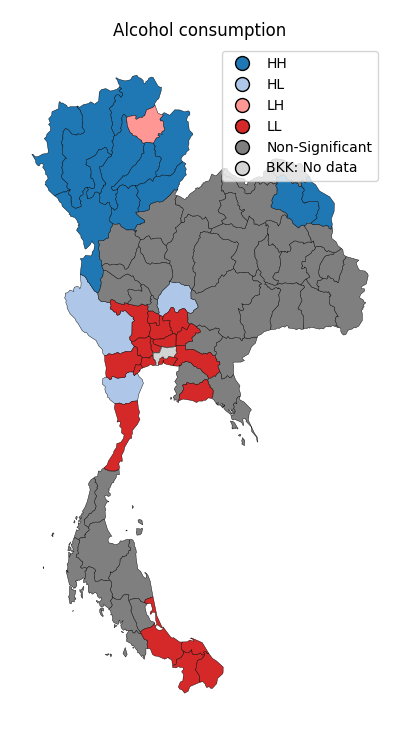}
	\end{minipage}
	\begin{minipage}[c]{0.24\textwidth}
		\includegraphics[width=\textwidth]{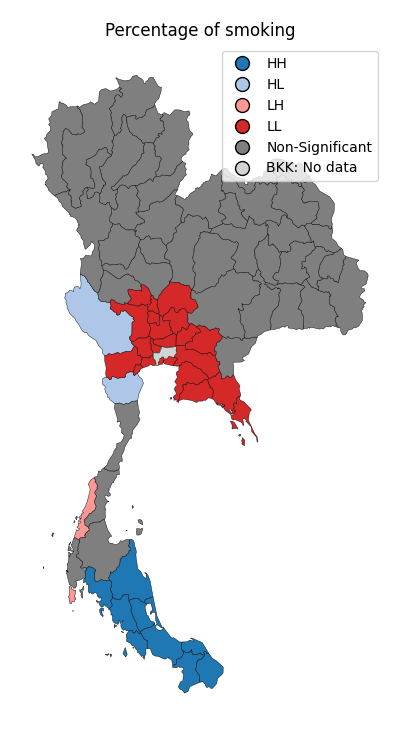}
	\end{minipage}
	\caption{Moran's clusters for the considered poverty factors.}
	\label{fig:AllVariablesMoranCluster}
\end{figure}

In \Cref{fig:AllVariablesMoranCluster} we present the Moran's clusters for the considered variables. We find a positive value for the Moran's I statistic for all the variables, which are reported in \Cref{tab:VariablesAspects}, and reject the hypothesis of no spatial structure with a $p$-value of 0.001.

Overall, we can detect five different clusters, with the following characteristics:

\begin{itemize}
\item \textbf{Bangkok Metropolitan Area and Eastern Thailand}: Around Bangkok and, in less extend, Eastern Thailand, we detect a cluster that overlaps for several variables, being a region with high level of education, low percentage of households in debt, as well as low percentage of alcohol consumption and smoking.
\item \textbf{Northeastern Thailand}: This region forms a cluster with low inequality (according to the monthly income ratio 20:20), a small number of households without savings, but a large percentage of households in debt.
\item \textbf{Northern Thailand}: As Northeastern Thailand, this region forms a cluster with a small number of households without savings, but with a low amount for such savings, it also depicts a cluster with high levels of alcohol consumption.
\item \textbf{Southern Thailand}: At Southern Thailand we detect a cluster with a high percentage of households without savings, but with low inequality (according with the Gini index), it appears to be a region with low levels of alcohol consumption but high percentage of smoking.
\item \textbf{Western Thailand}: This cluster is characterized by high levels of inequality (according to the monthly income ratio 20:20).
\end{itemize}

\subsection{Moran cluster for the first principal component}

While the previous analysis can help us to identify clusters based on their poverty dynamics, it has the disadvantage of considering only one variable at a time, which does not capture the possible interactions between them. To consider the multidimensional nature of poverty from a more integrated perspective, we perform a principal component analysis over the standardized variables. In \Cref{tab:VariancePca} we present the cumulative percentage of the variance explained by the principal components.

\begin{table}
	\centering
	\begin{tabular}{|ccccccccc|}
	\hline
	PC$_1$ & PC$_2$ & PC$_3$ & PC$_4$ & PC$_5$ & PC$_6$ & PC$_7$ & PC$_8$ & PC$_9$ \\
	\hline
	38\% & 61\% & 74\% & 86\% & 90\% & 94\% & 97\% & 98\% & 100\% \\
	\hline
	\end{tabular}
	\caption{Cummulative percentage of variance explained by the principal components.}
	\label{tab:VariancePca}
\end{table}

\begin{figure}
	\centering
		\begin{minipage}[c]{0.33\textwidth}
		\includegraphics[width=\textwidth]{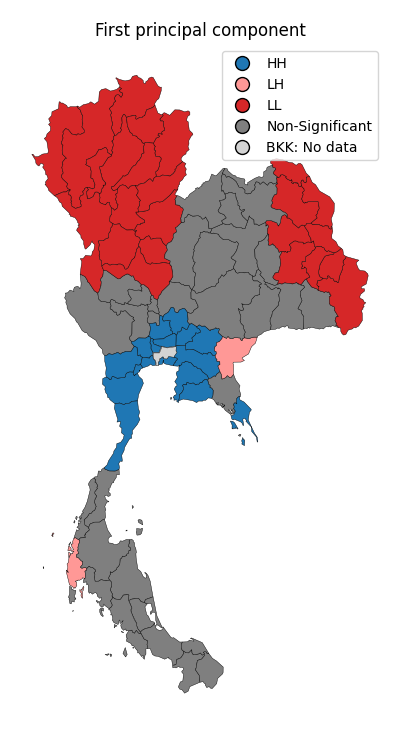}
	\end{minipage}
	\begin{minipage}[c]{0.33\textwidth}
		\includegraphics[width=\textwidth]{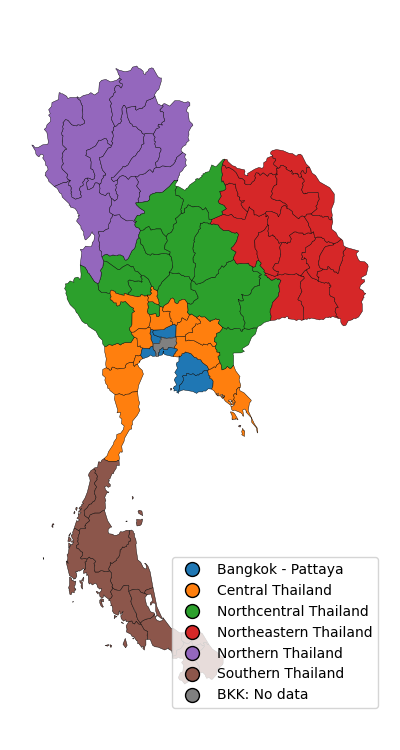}
	\end{minipage}
	\caption{Left: Moran's clusters using the first principal component. Right: Proposed regions using agglomerative hierarchical clustering, adding the spatial constraint.  It is interesting to notice that Northern Thailand and Northeastern Thailand correspond approximately with two of the Moran’s clusters for the first principal component. While Bagkok-Pattaya and Central Thailand form, approximately, the third cluster detected with the first principal component.}
	\label{fig:PcaMoranClusterAndProposedRegions}
\end{figure}

Then, we apply our previous approach on the first principal component, whose Moran's I statistic takes the value of 0.78, rejecting the hypothesis of no spatial structure with a $p$-value of 0.001. In this case, we detect three clusters, which are shown on the left side of \Cref{fig:PcaMoranClusterAndProposedRegions}. One of these clusters corresponds with most of the provinces of Northern Thailand. A second cluster is formed by the most eastern provinces of Northeastern Thailand. Lastly, we detect a third cluster formed by the provinces around Bangkok, Eastern Thailand and some provinces of Western Thailand. The first two clusters are characterized by a low value for the first principal component, while the third one is associated with a high value.

\begin{figure}
	\centering
	\includegraphics[width=\textwidth]{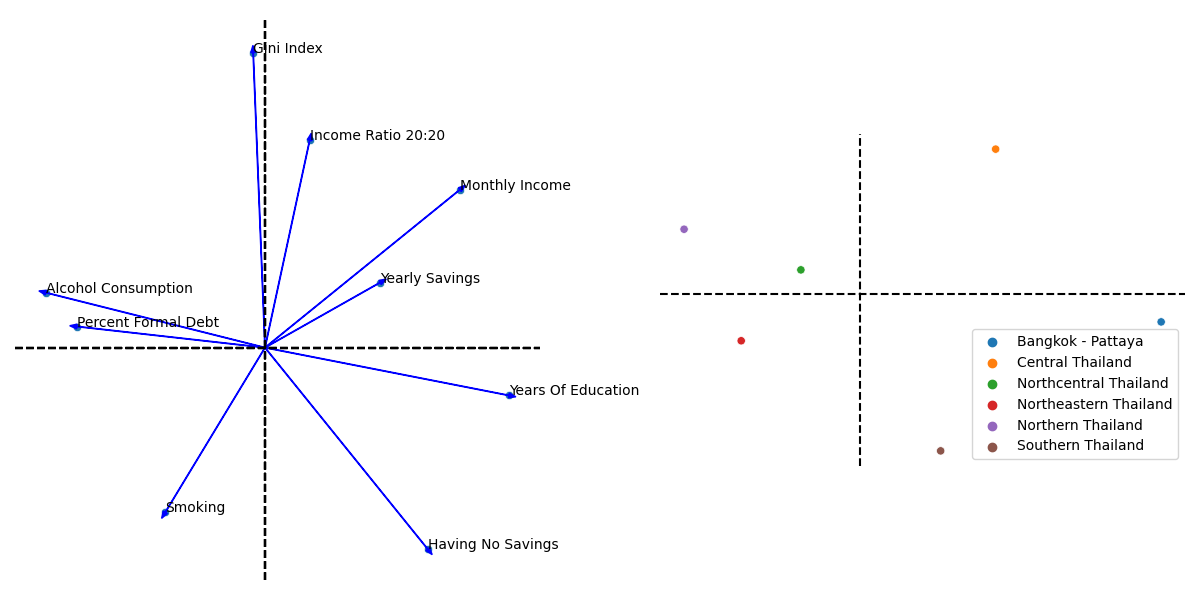}
	\caption{Biplot considering the first two principal components. Left: The loads of the variables are depicted as vectors in the cartesian coordinate form by the two first principal components. Right: We present the average value of the first two principal components for each one of the proposed regions.}
	\label{fig:Biplot}
\end{figure}

To get an interpretation for the clusters detected through the principal components, in \Cref{tab:PcaLoads} we present the loads of the variables for the first two principal components. These loads are presented graphically on the left side of the biplot in \Cref{fig:Biplot}.

\begin{table}
	\centering
	\begin{tabular}{|l|c|c|}
	\hline
	\textbf{Variable} & \textbf{PC$_1$} & \textbf{PC$_2$} \\
	\hline
	Years of education & 0.51 & -0.10 \\ 
	Monthly income & 0.40 & 0.33 \\ 
	Yearly savings & 0.24 & 0.13 \\ 
	Percentage of households without savings & 0.34 & -0.42 \\ 
	Monthly income ratio 20:20 & 0.09 & 0.43 \\ 
	Gini index & -0.02 & 0.61 \\ 
	Percentage of households with formal debt & -0.39 & 0.04 \\ 
	Alcohol consumption & -0.45 & 0.11 \\ 
	Smoking & -0.21 &-0.34 \\ 
	\hline 
	\end{tabular}
	\caption{Loads of the poverty factors for the first two principal components.}
	\label{tab:PcaLoads}
\end{table}

The first principal component can be interpreted mostly as a comparison between variables with an intrinsic positive aspect such as monthly income, yearly savings, and years of education; and variables with an intrinsic negative aspect such as alcohol consumption, percentage of formal debt, and smoking. Thus, generally speaking, the larger the value of the first principal component for a province, the more it is related with positive aspects, while the lower its value, the more it is related with negative aspects.

Therefore, we can conclude that the first two clusters identified through the first principal component, and the Moran's cluster approach, are more correlated with negative aspects associated to the multidimensional nature of poverty. Meanwhile, the third cluster, corresponding with the provinces around Bangkok, corresponds to a region more correlated with positive aspects. This helps to guide policy makers to identify regions that need more attention and to establish the high-priority aspects to mitigate.

\subsection{Regionalization}
\label{sec:Regionalization}

While Moran's clusters have worked so far to detect and characterized regions of Thailand based on their poverty dynamics, they present fragmented clusters, with several provinces not being assigned to any group. To ensure that clusters are not spatially fragmented and that every province is assigned to some cluster, we turn into regionalization.

\begin{figure}
	\centering
	\begin{minipage}[c]{0.45\textwidth}
		\includegraphics[width=\textwidth]{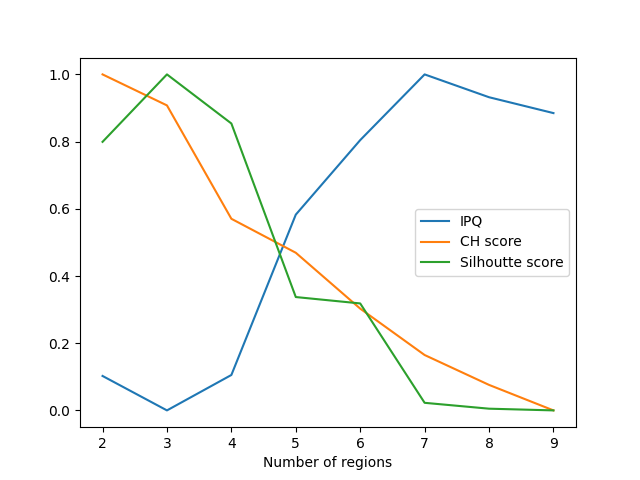}
	\end{minipage}
	\begin{minipage}[c]{0.45\textwidth}
		\includegraphics[width=\textwidth]{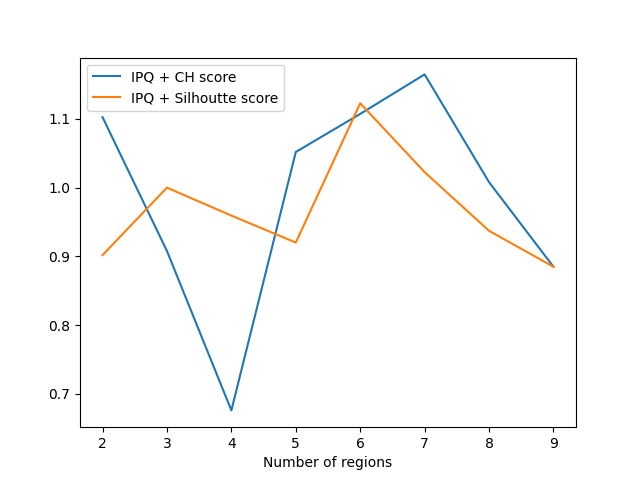}
	\end{minipage}
	\caption{Left: We present the $IPQ$ score, the silhoutte score and the Calinski-Harabasz score as functions of the number of regions considered, we have normalized these scores. Right: We consider the $IPQ$ plus each one of the feature coherence score to account for both, the spatial coherence and the goodness-of-fit.}
	\label{fig:GeographicFeatureCoherence}
\end{figure}

We consider an agglomerative hierarchical clustering, considering the ward linkage and adding the spatial constraint given by the matrix of spatial weights. The average value over the regions for the isoperimetric quotient, the silhoutte score, and the Calinski-Harabasz score are presented on the left side of \Cref{fig:GeographicFeatureCoherence}, against the number of regions, which vary from 2 to 9 regions, normalizing the metrics, so the maximum takes the value of 1 while the minimum takes the value of 0. We observe that while the $IPQ$ tends to increase with the number of regions, the silhoutte score and the Calinski-Harabasz score tend to decrease. Thus, it is necessary to make a trade-off between the geographic and the feature coherence.

As a simple way to count at the same time for the geographical and feature coherence, we simply sum the geographical metric of coherence ($IPQ$) and each one of the goodness-of-fit metrics. These new metrics are presented on the right side of \Cref{fig:GeographicFeatureCoherence}. We observe that, according to these new metrics, the optimum number of regions is around 6 or 7.

Through trial and error, we found 6 regions to be a good choice for the number of regions in Thailand, considering their capability to explain the dynamics of each region. We have named these regions as: Bangkok-Pattaya, Central Thailand, Northcentral Thailand, Northeastern Thailand, Northern Thailand, and Southern Thailand, which are shown on the right side of \Cref{fig:PcaMoranClusterAndProposedRegions}. It is interesting to notice that Northern Thailand and Northeastern Thailand correspond approximately with two of the Moran's clusters for the first principal component. While Bagkok-Pattaya and Central Thailand form, approximately, the third cluster detected with the first principal component.

On the right side of \Cref{fig:Biplot} we present the average value of the two principal components for these regions. We observe that the Bangkok-Pattaya region, Central Thailand, and in less extend Southern Thailand correspond with the regions with the largest values for the first principal component. And, hence the regions less associated with negative aspects of the poverty dynamics. However, if we also consider the second principal component, we can detect that positive values for this component might be more associated with inequality within the region, while a negative value might be due to a high percentage of households without savings and/or high levels of smoking. Thus, we can identify Central Thailand as a region with high levels of inequality, and Southern Thailand as a region with a high percentage of households without savings and high levels of smoking.

On the other hand, Northern Thailand, Northeastern Thailand, and in less extend Northcentral Thailand correspond with the regions with the lowest values for the first principal component, all of them with a negative value for their mean. And, thus with the regions that require more attention.

\begin{figure}
	\centering
	\includegraphics[width=0.85\textwidth]{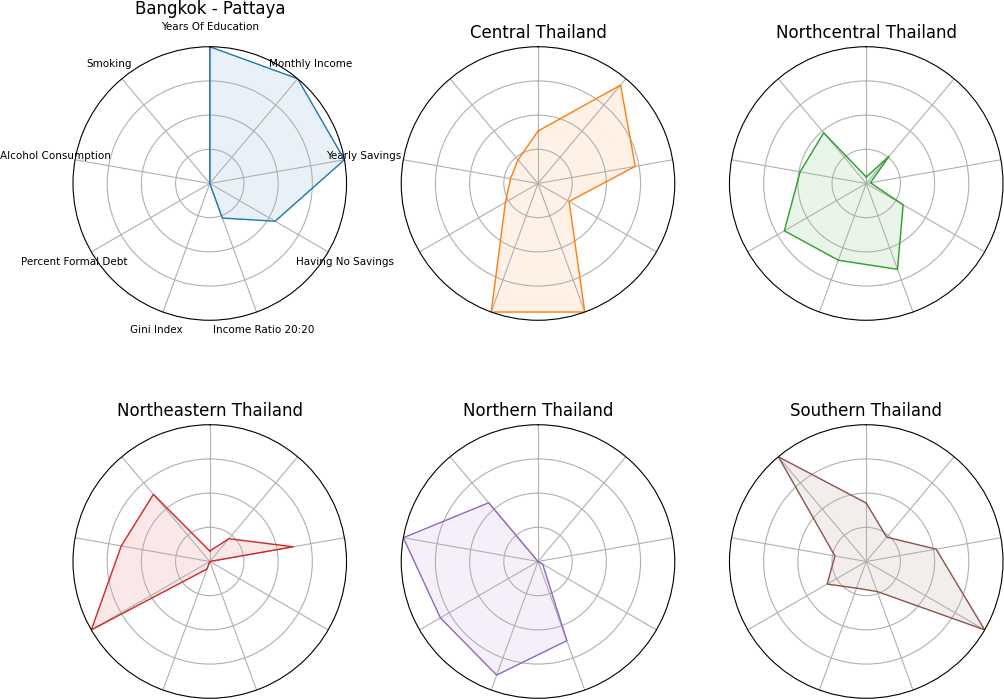}
	\caption{Profile for each proposed region, all the variables have been normalized.}
	\label{fig:RadarChart}
\end{figure}

To complement the profile of each region, in \Cref{fig:RadarChart} we present the radar chart for each one of the identified regions, normalizing the variables, providing a whole profile which can help us to identify the principal issues faced for each region.

\begin{itemize}
\item \textbf{Bangkok-Pattaya}: This region corresponds with the wealthiest region of Thailand, principally associated with high levels of education, income and savings. And, at the same time, keeping one the lowest levels for the rest of the variables, which posse an inherent negative aspect.
\item \textbf{Central Thailand}: While it presents similar values to Bangkok-Pattaya for the monthly income and the amount of savings, important differences appear for the years of education, with a dramatic decrease in the levels of education. We also find Central Thailand to be the region with the largest value for the income ratio 20:20 and the Gini index, thus we can identify this region as a buffer zone of transition from the wealthy region of Bagkok-Pattaya and the rest of the country.
\item \textbf{Northcentral Thailand}: This region does not seem to be associated with any issue in particular, so we can interpret it as a second transition region from Central Thailand to Northern and Northeastern Thailand.
\item \textbf{Northeastern Thailand}: This region is principally characterized by high levels of debt and in less extend to alcohol consumption and smoking.
\item \textbf{Northern Thailand}: This region is principally associated with high levels of alcohol consumption, a high percentage of households in debt, and inequality, according to the Gini index and, in less extend, to the income ratio 20:20.
\item \textbf{Southern Thailand}: This region is associated with the highest levels of smoking and the largest percentage of households without savings.
\end{itemize}

We can also observe that Northcentral, Northeastern, Northern and Southern Thailand present pretty low levels of education, income and amount of savings contrasted with Bagkok-Pattaya and Central Thailand.

\subsection{Bayesian hierarchical regression}

Following the approach proposed in ~\cite{gomez2024income}, we fit a hierarchical regression model to estimate the impact of years of education into the income for the proposed regions. In this regression framework, the monthly income for province $i$ belonging to the region $j$, $Y_{ij}$ is modeled as \[Y_{ij}|\alpha_j,\beta_j,\sigma_j \sim \textsf{Laplace}(\alpha_j + \beta_j (X_{ij}-\bar{X}_{\cdot j}),\sigma_j),\]
where $X_{ij}$ is the average years of education in the province $i$, and \[\bar{X}_{\cdot j}=\frac{\sum_{i=1}^{n_j} X_{ij}}{n_j}.\] Adding a hierarchical structure into $\alpha_j$ and $\beta_j$ to model the national level.

\begin{figure}
	\centering
	\includegraphics[width=\textwidth]{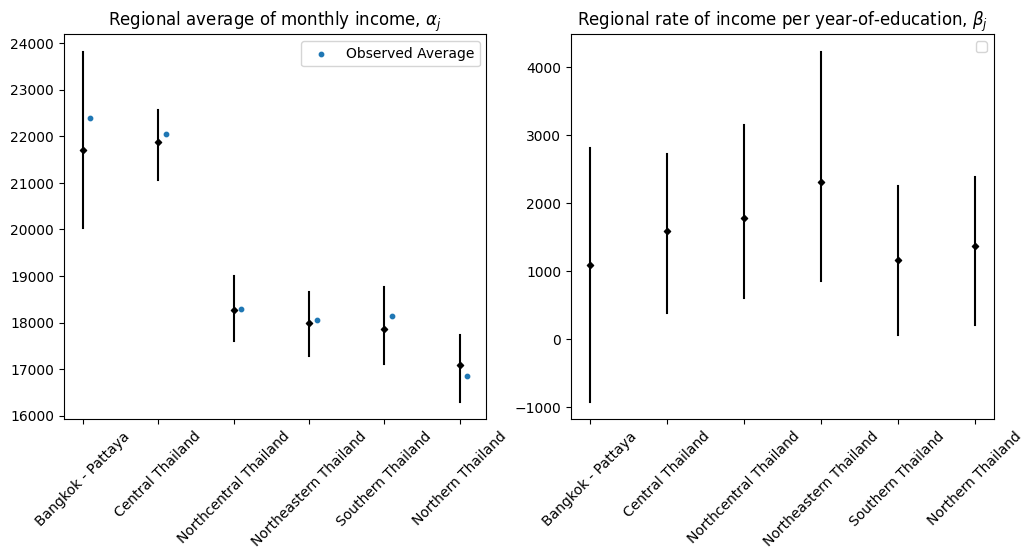}
	\caption{Credible intervals of 0.95 posterior probability for the regional average monthly income and the regional rate of income per year-of-education.}
	\label{fig:AlphaBetaIntervals}
\end{figure}

\begin{figure}
	\centering
	\begin{minipage}[c]{0.29\textwidth}
		\includegraphics[width=\textwidth]{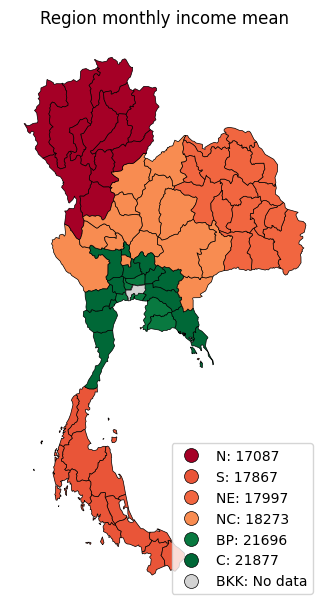}
	\end{minipage}
	\begin{minipage}[c]{0.29\textwidth}
		\includegraphics[width=\textwidth]{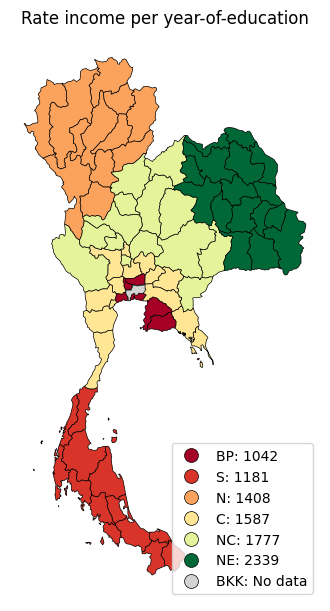}
	\end{minipage}
	\caption{Posterior mean for the regional average of monthly income and the rate of income per year-of-education.}
	\label{fig:AlphaBetaMap}
\end{figure}

\begin{figure}
	\centering
	\includegraphics[width=\textwidth]{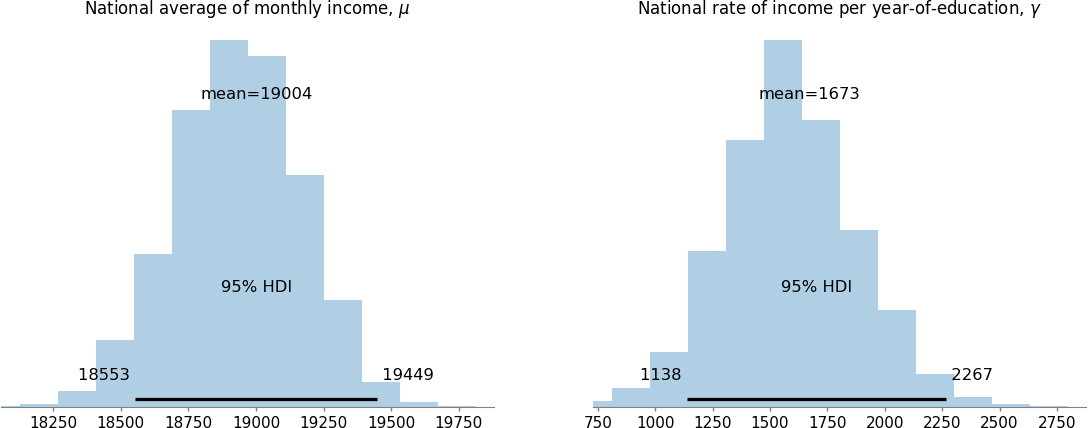}
	\caption{Posterior mean for the national average of monthly income and the rate of income per year-of-education.}
	\label{fig:MuGammaDistributions}
\end{figure}

In \Cref{fig:AlphaBetaIntervals} we present the credible intervals of 0.95 posterior probability for the regional average monthly income and the regional rate of income per year-of-education. Bangkok-Pattaya and Central Thailand show a similar income, being far higher than the income for the rest of the regions. Northcentral, Northeastern and Southern Thailand also show a similar income between them. While Northern Thailand presents the lowest income between all the regions. For the rate of income per year-of-education, we observe a large overlapping for most of the regions, which might indicate that all the regions share a common rate. The posterior mean for each one of these parameters are also mapped in \Cref{fig:AlphaBetaMap}. In \Cref{fig:MuGammaDistributions} we present the posterior distributions of the national average monthly income and the national average rate of income per year-of-education. Getting similar results for the national level as those presented in ~\cite{gomez2024income}.

\begin{figure}
	\centering
	\includegraphics[width=\textwidth]{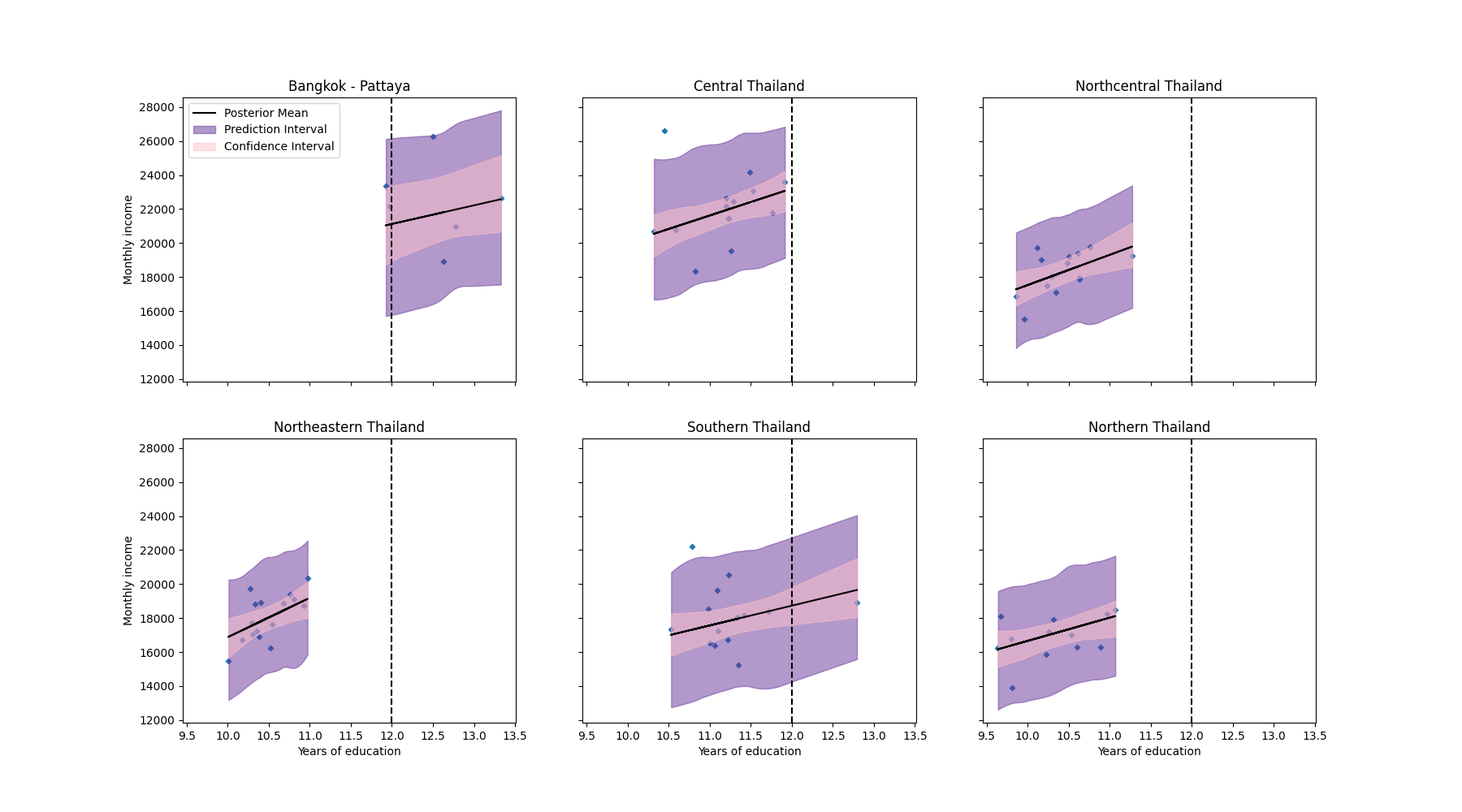}
	\caption{Regression fitted for each one of the regions. For reference, we have added a vertical dashed line at 12 years of education, which corresponds (approx.) with a complete senior high school. The national average is of 10.87
years.}
	\label{fig:HierarchicalRegression}
\end{figure}

Finally, in \Cref{fig:HierarchicalRegression} we present the regression fitted for each one of the regions. For reference, we have added a vertical dashed line at 12 years of education, which corresponds (approx.) with a complete senior high school. We observe that Bangkok-Pattaya is the only region whose average years of education is above the 12 years threshold, with most of the provinces in the region above the threshold, with the only exceptions of Chonburi and Rayong which are a little below the threshold with 11.98 and 11.93 years, respectively. On the other hand, except for Phuket with 12.79 years of education on average, all the provinces that do not belong to the Bangkok-Pattaya region have an average years of education below 12 years, while the national average is of 10.87 years. In \Cref{tab:EducationForRegion} we present the average years of education for each region.

\begin{table}
	\centering
	\begin{tabular}{|l|c|}
	\hline
	\textbf{Region} & \textbf{Years of education} \\
	\hline
	Bangkok-Pattaya & 12.52 \\
	Southern Thailand & 11.26 \\
	Central Thailand & 11.16 \\
	Northeastern Thailand & 10.49 \\
	Northcentral Thailand & 10.42 \\
	Northern Thailand & 10.32 \\
	\hline
	Thailand & 10.87 \\
	\hline
	\end{tabular}
	\caption{Average years of education for each region.}
	\label{tab:EducationForRegion}
\end{table}

\begin{figure}
	\centering
	\includegraphics[width=0.75\textwidth]{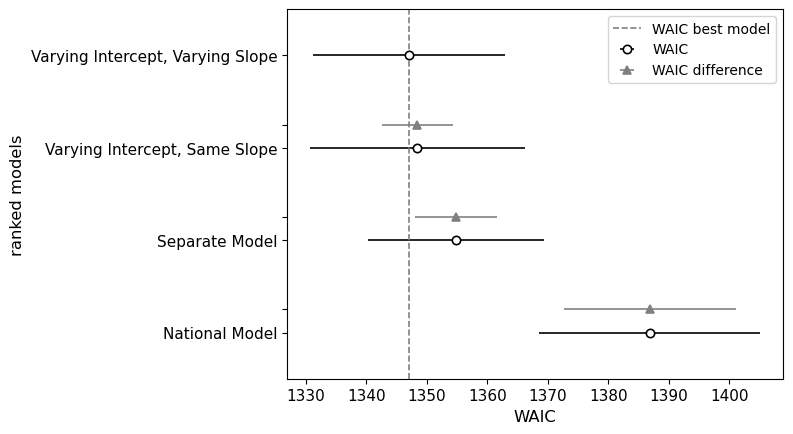}
	\caption{WAIC for the four Bayesian hierarchical models proposed in ~\cite{gomez2024income}.}
	\label{fig:WAIC}
\end{figure}

Due to the large overlap of the credible intervals for the regional rate of income per year-of-education, as it is shown on the right side of \Cref{fig:AlphaBetaIntervals}, we might better opt for a hierarchical model that considers one common rate of income per year-of-education. For this purpose, we also follow the model presented by ~\cite{gomez2024income} with a common parameter $\beta$ for all the regions. For completion, we have fitted separate independent models for each region and a single common model for all regions. For comparison purposes, in \Cref{fig:WAIC} we present the Widely Applicable Information criterion (WAIC) for these four models. We observe that independent models and a single common model for all regions present a significantly larger WAIC that the hierarchical models, and thus should not be considered anymore.

In \Cref{tab:ModelDiffAlphaDiffBeta,tab:ModelDiffAlphaSameBeta} we present the estimated values for the monthly income and the rate income per year-of-education at a regional and a national levels for the hierarchical models considering different slopes, and a common slope, respectively. Together with the posterior mean of the parameters, we present an interval of 0.95 posterior probability. While we obtain mostly the same estimates for the monthly income, and the rate income per year-of-education at a national level, it is worthy to notice that the posterior interval is slightly smaller when we consider a common slope. On the other hand, when we consider different slopes for the regions, the posterior interval for some of these slopes include negative values, which is very unlikely. Thus, a common slope for all regions seems to be a model that can explain better our phenomenon.

\begin{table}
	\centering
	\begin{tabular}{|l|c|c|}
	\hline
	\textbf{Region} & \textbf{Monthly income mean} & \textbf{Rate income per year-of-education} \\
	\hline
	National level & 19004; (18553, 19449) & 1673; (1138, 2267) \\
	\hline
	Bangkok - Pattaya & 21683; (19927, 23550) & 1042; (-1062, 2676) \\
	Central Thailand & 21854; (21090, 22520) & 1587; (399, 2803) \\
	Northcentral Thailand & 18261; (17522, 18950) & 1777; (559, 3164) \\
	Northeastern Thailand & 18057; (17380, 18700) & 2339; (947, 4300) \\
	Northern Thailand & 17859; (17166, 18682) & 1181; (47, 2262) \\
	Southern Thailand & 17078; (16316; 17727) & 1408; (156, 2445) \\
	\hline
	\end{tabular}
	\caption{Posterior mean and interval of 0.95 posterior probability for the hierarchical model that consider different intercepts and slopes. Most of the intervals for the slope present a large overlap, which might indicate a common slope for all the regions.}
	\label{tab:ModelDiffAlphaDiffBeta}
\end{table}

\begin{table}
	\centering
	\begin{tabular}{|l|c|c|}
	\hline
	\textbf{Region} & \textbf{Monthly income mean} & \textbf{Rate income per year-of-education} \\
	\hline
	National level & 19015; (18554, 19485) & 1545; (982, 2059) \\
	\hline
	Bangkok - Pattaya & 21779; (19828, 23567) & \\
	Central Thailand & 21852; (20804, 22892) & \\
	Northcentral Thailand & 18310; (17638, 18982) & \\
	Northeastern Thailand & 18084; (17441, 18754) & \\
	Northern Thailand & 18194; (17242, 19167) & \\
	Southern Thailand & 16943; (16157; 17722) & \\
	\hline
	\end{tabular}
	\caption{Posterior mean and interval of 0.95 posterior probability for the hierarchical model that consider different intercepts and a common slope for all the regions.}
	\label{tab:ModelDiffAlphaSameBeta}
\end{table}

\subsection{Spatial regression}

\begin{figure}
	\centering
	\begin{minipage}[c]{0.31\textwidth}
		\includegraphics[width=\textwidth]{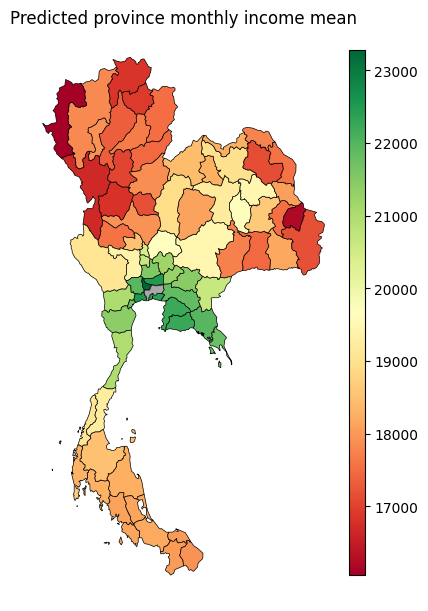}
	\end{minipage}
	\begin{minipage}[c]{0.29\textwidth}
		\includegraphics[width=\textwidth]{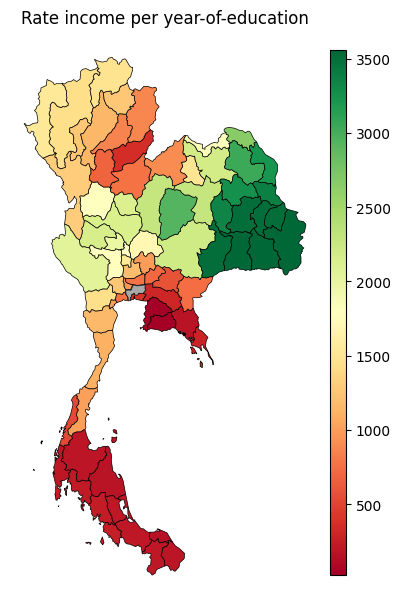}
	\end{minipage}
	\caption{Results of the GWR model. Left: we present the estimated monthly average income for each province, getting similar values to the observed ones. Right: we present the estimated rate income per year-of-education for each province.}
	\label{fig:GwrResults}
\end{figure}

Considering the spatial weights, we can fit a geographically weighted regression (GWR) model, explaining the province income $Y_i$ as a function of its average years of education in such province. On the left side of \Cref{fig:GwrResults} we present the estimated monthly average income for each province, getting similar values to the observed ones, presented in the first map of \Cref{fig:SpatialNetwork}. However, note that the most extreme observed values are diminished. On the right side of \Cref{fig:GwrResults} we present the estimated rate income per year-of-education for each province, it worth notice that there are no negative values for these estimations, even while such constriction was not included explicitly into the model.

\begin{figure}
	\centering
	\includegraphics[width=\textwidth]{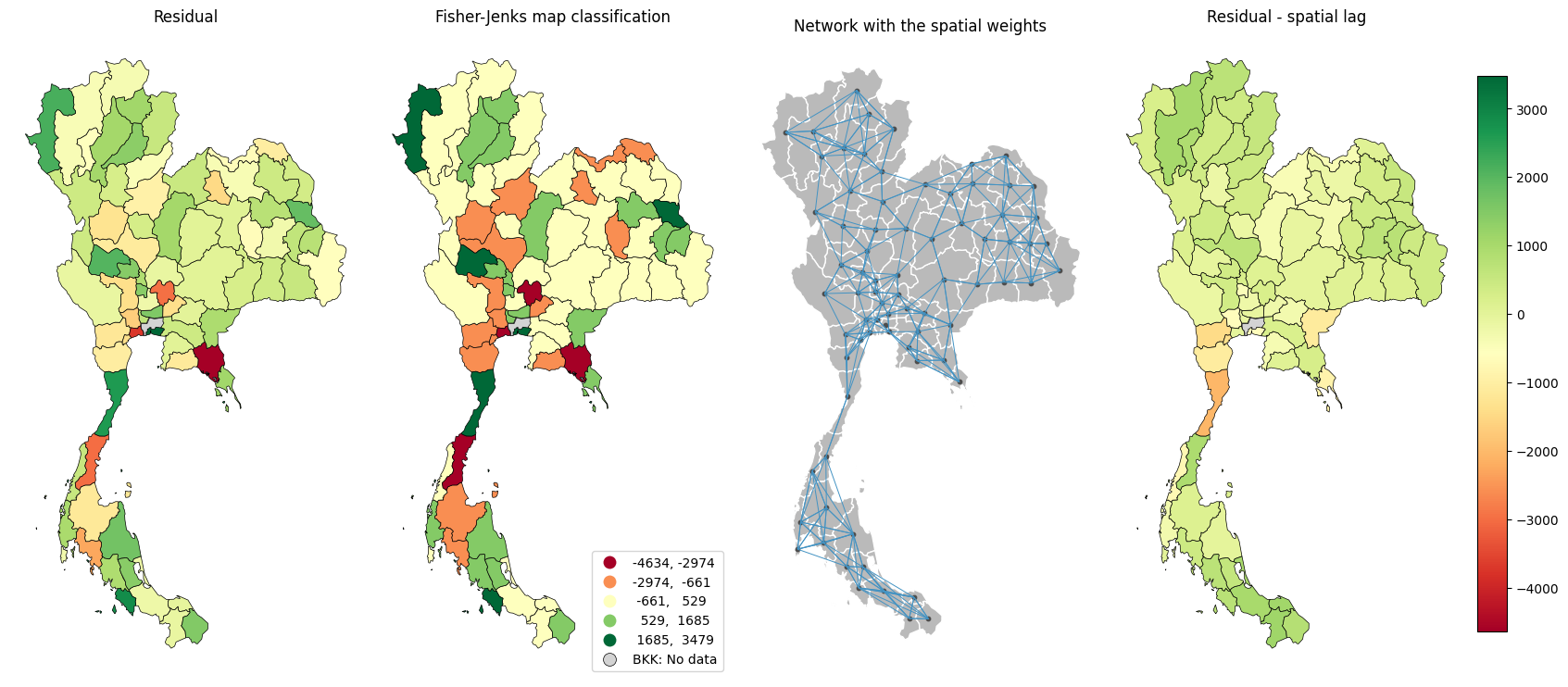}
	\caption{We map the residuals of the GWR model and their spatial lag. Note that the residuals do not present any geographical pattern.}
	\label{fig:GwrResiduals}
\end{figure}

\begin{figure}
	\centering
	\includegraphics[width=\textwidth]{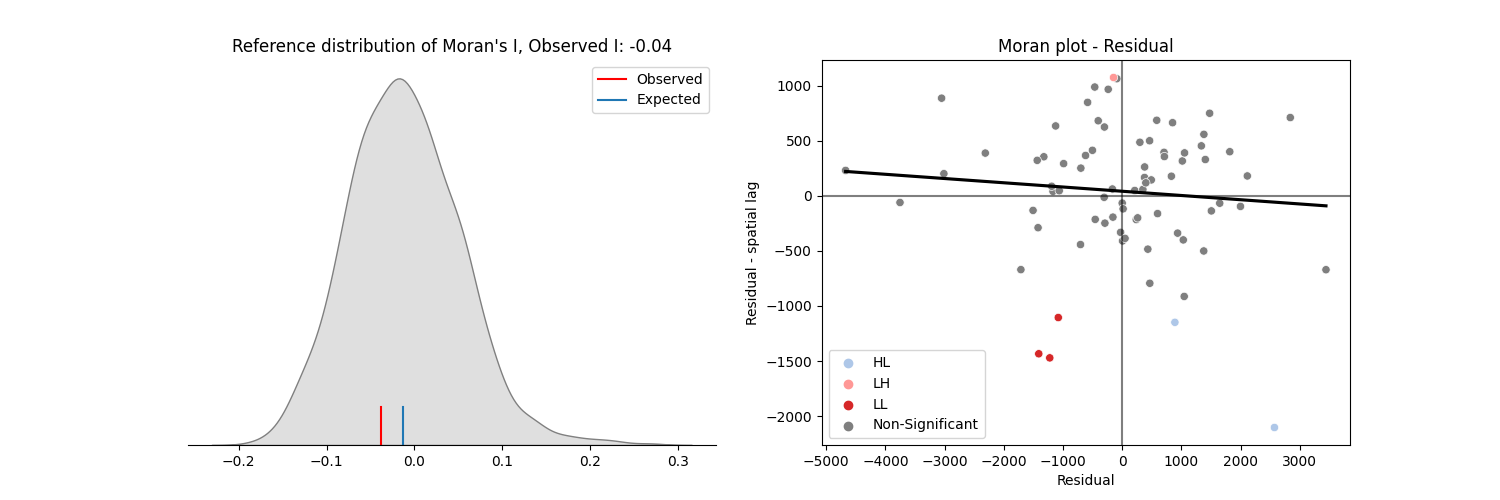}
	\caption{Reference distribution of the Moran's I for the residuals of the model, and their respective Moran's scatter plot. We do not reject the hypothesis of no spatial correlation.}
	\label{fig:GwrResidualsMoran}
\end{figure}

To evaluate our spatial regression model, we analyze its residuals. In \Cref{fig:GwrResiduals} we map the residuals of the model and their spatial lag. It is difficult to observe any geographical pattern in the residuals, being an indicative that the model can recover and estimate correctly the geographical pattern of the data. In \Cref{fig:GwrResidualsMoran}, we present the reference distribution of the Moran's I for the residuals of the model, with a $p$-value of 0.371 for the hypothesis of no spatial correlation, which is sufficiently large to not reject such hypothesis, being a desirable property of a well-fitted spatial regression model.

\section{Discussion and conclusion}
\label{sec:Discussion}

Since a single one-shirt-size policy is unable to take into account differences between all regions of Thailand and making a specific policy for each region is impossible due to the limitation of time and resources, finding an optimum regional configuration becomes necessary to implement the right policy and alleviate the poverty issues.

With this challenge in mind, throughout this work, we presented a statistical spatial analysis to inferred optima geographical regions. Usual clustering techniques such as the Fisher-Jenks algorithm do not take into account the spatial structure to inferred clusters, thus, they present the disadvantage of creating disconnected geographic clusters. Hence, a statistical analysis to inferred regions that consider the spatial structure becomes necessary.

The first step to inferred an optimum regional configuration is to create a network that reflects the connections between the provinces of Thailand. We observed that considering the five nearest neighbor provinces to build such network maintains the analysis local, while reproducing the geographical structure of Thailand. These spatial weights can be used to calculate a statistic of spatial autocorrelation for a variable of interest. We have used Moran's I statistic for this purpose. All poverty factors considered in this work presented a positive I statistic, rejecting the hypothesis of abscence of spatial structure, and suggesting the existence of clusters for provinces with similar poverty dynamics. To inferred such geographical clusters, we have used local Moran's I statistic and Moran's scatterplot. However, this is a univariate approach, and thus cannot take into account the multidimensional poverty phenomenon. To partially solve this problem, we first performed a PCA analysis, finding four well-distinguished Moran clusters when we considered the first principal component. However, an important drawback of Moran clusters is that they are created by provinces where the hypothesis of no-spatial structure is rejected, leaving several provinces without being assigned to any cluster.

To take into account the non-linear interactions between the variables and the spatial structure, we inferred regions trough an agglomerative hierarchical clustering technique. Considering the isoperimetric quotient as a geographic coherence metric, and the silhoutte score and the Calinski-Harabasz score as goodness-of-fit. We found six regions as an optimum number for explaining the multidimensional aspects of poverty, and created a profile for every region, which helps to identify regions that need more attention and to establish the high-priority aspects to mitigate in each one of them. This approach has the advantage of creating non-fragmented clusters, assigning every province to some region.

The inferred regions can then be incorporated into more complex models, such as Bayesian hierarchical regression models, explaining some variable of
interest in terms of other variables at a regional and national levels. Furthermore, using the spatial weights, it is possible to implement GWR models, which take into account the geographical structure. Using these approaches we were able inferred the effect of the education in the monthly income at a province, regional and national levels while considering the geographical structure of the country.

With all results above, we found that 1) Northern, Northeastern Thailand, and in less extend Northcentral Thailand are the regions that require more attention in the aspect of poverty issues, 2) Northcentral, Northeastern, Northern and Southern Thailand present dramatically low levels of education, income and amount of savings contrasted with large cities such as Bagkok-Pattaya and Central Thailand, and 3) Bangkok-Pattaya is the only region whose average years of education is above 12 years, which corresponds (approx.) with a complete senior high school.

In conclusion, the statistical spatial analysis presented throughout this work has the potential to inferred boundaries for regions in Thailand that can explain better the poverty dynamics, instead of the usual government administrative regions. The proposed regions maximize a trade-off between feature and geographical coherence. Such regions can be incorporated into more complex models, such as Bayesian hierarchical regression models, with the potential of assist the implementation of the right policy to alleviate the poverty phenomenon.

 
\section*{Note}
Codes to reproduce our results are available in \href{https://github.com/IrvingGomez/SpatialPovertyFactors}{https://github.com/IrvingGomez/SpatialPovertyFactors}

\section*{Acknowledgement}
Throughout this article we have used extensively the \texttt{libpysal} library ~\cite{pysal2007}. The authors would like to give a special thanks to Sergio Rey who patiently helped with some technical aspects of the library. The authors also would like to give a special thanks to Dr. Anon Plangprasopchok and Ms. Kittiya Ku-kiattikun who managed the data, cleaned it, and provided insight about the data for us. 

\bibliographystyle{unsrt}


\end{document}